\documentclass[sigconf,nonacm]{acmart}

\usepackage{amsmath}
\usepackage{algorithm}
\usepackage[noend]{algpseudocode}
\usepackage{multicol}
\usepackage{graphicx}
\usepackage{subcaption}
\usepackage{wrapfig}
\usepackage{paralist}
\usepackage{xspace}
\usepackage{enumitem}
\usepackage{mathtools}
\usepackage{algorithm}
\usepackage[noend]{algpseudocode}
\usepackage{booktabs}
\usepackage{wrapfig}
\usepackage{balance}
\usepackage[font=small,skip=1pt]{caption}
\usepackage[first=0,last=9]{lcg}
\usepackage{color, colortbl}
\usepackage{float}
\usepackage{comment}
\usepackage[capitalise]{cleveref}
\usepackage[subtle]{savetrees}
\setlist{topsep=0pt, leftmargin=*}

\newcommand{\name}{TailorSQL\xspace}
\newcommand{\NewPara}[1]{\vspace{4pt}\noindent{\bf #1}}

\author[K. Vaidya, J. Ding, S. Kosak, D. Kernert, C. Lei, X. Qin, A. Tripathy, R. Balan, B. Narayanaswamy, T. Kraska]
{
    Kapil Vaidya$^{1*}$, Jialin Ding$^2$, Sebastian Kosak$^{3*}$, David Kernert$^{4*}$, Chuan Lei$^2$, Xiao Qin$^2$, Abhinav Tripathy$^2$, Ramesh Balan$^2$, Balakrishnan Narayanaswamy$^2$, Tim Kraska$^2$
}
\affiliation{%
  \institution{$^1$Parallel Web Systems\quad$^2$Amazon Web Services \quad$^3$Technical University of Munich\quad$^4$STACKIT}
  \city{}
  \country{}
}
\thanks{$^*$Work performed while employed at AWS}

\begin{document}
\title{TailorSQL: An NL2SQL System Tailored to Your Query Workload}

\begin{abstract}

NL2SQL (natural language to SQL) translates natural language questions into SQL queries, thereby making structured data accessible to non-technical users, serving as the foundation for intelligent data applications. State-of-the-art NL2SQL techniques typically perform translation by retrieving database-specific information, such as the database schema, and invoking a pre-trained large language model (LLM) using the question and retrieved information to generate the SQL query.

However, existing NL2SQL techniques miss a key opportunity which is present in real-world settings: NL2SQL is typically applied on existing databases which have already served many SQL queries in the past. The past query workload implicitly contains information which is helpful for accurate NL2SQL translation and is not apparent from the database schema alone, such as common join paths and the semantics of obscurely-named tables and columns. We introduce TailorSQL, a NL2SQL system that takes advantage of information in the past query workload to improve both the accuracy and latency of translating natural language questions into SQL. By specializing to a given workload, TailorSQL achieves up to 2$\times$ improvement in execution accuracy on standardized benchmarks.
\end{abstract}

\maketitle

\section{Introduction}
\label{sec:introduction}
NL2SQL translates natural language questions into SQL queries, making data analysis accessible to non-technical users and serving as a foundation for intelligent data applications like smart dashboards and visualizations.
For instance, a business owner can simply ask, ``What were last month's total sales by product?'' and the data application can use the NL2SQL system to generate the corresponding SQL, retrieve the relevant data, and summarize the results. Such possibilities make NL2SQL an important tool to explore.

NL2SQL tools are increasingly being integrated into widely-used commercial database systems~\cite{microsoft-nl2sql,redshift-nl2sql,snowflake-nl2sql}. Usage of NL2SQL tools typically exhibit two behaviors. First, NL2SQL-generated queries are only a small fraction of the overall query workload that is executed on a database; most SQL queries are still either hand-written or machine-generated according to application logic. Second, users of NL2SQL tools care not only about the accuracy of the output SQL query, but also the latency of invoking the NL2SQL tool.

Recent improvements in NL2SQL technology has been propelled by the advent of large language models (LLMs). What sets LLMs apart from the previous language models is their capacity to generalize to unseen tasks. LLMs achieve this through in-context learning, where examples of the task completion and information relevant to the task are provided within their context. Using this context, the LLM generates the desired output.
Capitalizing on these capabilities, LLMs consistently excel in various tasks, including NL2SQL, where they currently lead on two well-known NL2SQL benchmarks: Spider~\cite{yu2019spider} and BIRD~\cite{li2023llmBird}.

Since NL2SQL aims to output SQL queries that can execute on the user's database, one key step in the NL2SQL pipeline is acquiring database-specific information (e.g., the names of tables which the SQL query should reference) to provide in the LLM's context. Some commercial NL2SQL tools ask the user to manually specify certain information, such as the tables and views to be used in the SQL query~\cite{microsoft-nl2sql} or the literals to use in filters~\cite{snowflake-nl2sql}. On the other hand, NL2SQL techniques proposed in the research literature typically aim to automatically retrieve the relevant database-specific information without any human intervention~\cite{pourreza2023dinsql,wang2023mac,gao2023texttosql}.

However, one source of database-specific information is consistently ignored or underutilized by existing NL2SQL tools: the logs of past SQL queries which have executed on the database. These logs typically contain queries originating from various applications, including machine learning pipelines, ETL processes, and reporting tools, rather than being limited to those generated by NL2SQL interfaces. Databases often have repetitive workloads with similar queries in historical logs due to overlapping analyses driven by consistent business goals and standard reports. This leads to recurrences of similar patterns across queries, such as frequently-used join paths. As a result, SQL queries generated by NL2SQL systems often match or share components with logged queries.

\begin{figure*}[t]
    \begin{center}
        \includegraphics[width=\linewidth]{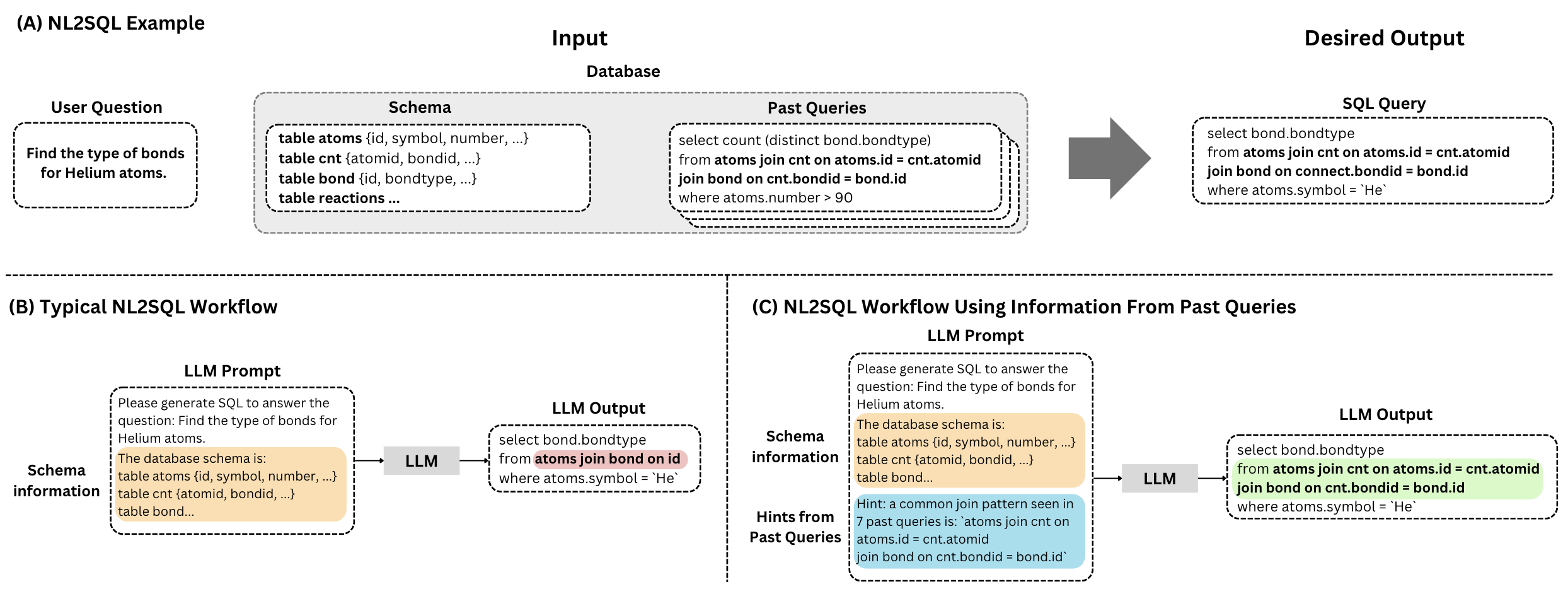}
    \end{center}
    \caption{\textbf{(A)} The input to NL2SQL is a user question over a database which has a certain schema and past query workload, and the desired output is a SQL query. \textbf{(A)} The typical NL2SQL workflow will invoke the LLM with a prompt that includes the user question and the database's schema, which may not include enough information to produce an accurate SQL query. \textbf{(C)} Hints extracted from past queries help the LLM produce more accurate queries.}
    \label{fig:main_example}
\end{figure*}

\NewPara{An Illustrative Example.}
We illustrate the potential benefit of using past query logs for NL2SQL with an example which is derived from the BIRD benchmark~\cite{li2023llmBird}. \cref{fig:main_example}A depicts a chemistry database containing tables named \texttt{atoms}, \texttt{cnt}, and \texttt{bond}, among others. Notably, the cryptically-named \texttt{cnt} table is meant to ``connect'' the \texttt{atoms} and \texttt{bond} tables via their respective \texttt{id} fields. Assume that a number of SQL queries have already been run on this chemistry database, not necessarily generated from the NL2SQL interface. Given the user's natural language question, an accurate NL2SQL system should produce the desired output SQL query, which joins \texttt{atoms} and \texttt{bond} via the \texttt{cnt} table.

A typical NL2SQL pipeline (\cref{fig:main_example}B) will invoke an LLM with a prompt that includes the user's question and schema information from the database. However, due to the cryptic name of the \texttt{cnt} table, the LLM might erroneously choose an incorrect join path: in this example, it joins the \texttt{atoms} and \texttt{bond} tables on their respective \texttt{id} columns. Figuring out the correct join path based solely on table schemas is challenging for the LLM, especially if there is no join path information and tables have no conclusive names.

To mitigate this lack of information, we can take advantage of past queries to augment the information provided in the prompt (\cref{fig:main_example}C): suppose that the join path between the \texttt{atoms}, \texttt{cnt}, and \texttt{bonds} tables has been observed in multiple past queries, such as the query shown in \cref{fig:main_example}A. We incorporate this common join path as a \textit{hint} in the LLM prompt, which guides the LLM to generate the correct SQL query. In this example, past queries provide information about common query patterns, which implicitly indicate the appropriate usage of schema items which may otherwise be unclear from inspecting the database schema alone. Note that hints are not directives, i.e., they do not force the LLM to act a certain way, but rather provide the LLM with useful information which the LLM is allowed to ignore.
\\
\\
Following this intuition, we introduce \name, a NL2SQL system that tailors itself to a given query workload by harnessing the information that is implicitly stored in past queries. \name improves both the accuracy and latency of NL2SQL with two core ideas.

First, \name performs an offline analysis of the past query workload and extracts \textit{hints} from past queries (e.g., the hint about the common join path in \cref{fig:main_example}). Hints provide useful information for accurate NL2SQL translation which is missing from schema information alone. Individual hints are stored as text-based \textit{documents} and can be retrieved into the LLM context if relevant for a given question. \name also stores schema information in documents.

Second, \name introduces a workload-specialized framework for retrieving the relevant documents for a given user question, while ignoring irrelevant documents. Retrieval frameworks (see \cref{sec:preliminaries}) are often necessary for real-world NL2SQL pipelines because LLMs are subject to a context limit, which is the maximum amount of input text the LLM can process. Database schemas often comprise thousands of tables and including all of them in the limited prompt context becomes infeasible. Even in cases where the entire schema fits into the prompt context, it is desirable to avoid adding irrelevant information to an LLM's prompt, since doing so will incur higher latency and cost when invoking the LLM. A typical NL2SQL retrieval framework maps user questions and documents into an embedding space, ensuring the proximity of questions to relevant documents; then for a given question, the framework retrieves documents in order of decreasing embedding similarity until the LLM's context is filled. \name's retrieval framework improves upon the typical framework by using the past query workload to create fine-tuned document embeddings, which improves its ability to retrieve relevant documents while avoiding irrelevant documents. Furthermore, instead of using a single context size limit for all documents, \name performs an analysis over past queries to allocate a separate context limit for each class of document (i.e., schema vs. hint documents) in order to mitigate class imbalances when performing document retrieval.

Finally, given that \name specializes to a specific workload, it may perform poorly if the workload distribution changes in the future. To mitigate such performance regressions, we introduce an \textit{abstention policy}, which \name uses to dynamically decide whether to abstain from using query hints in the NL2SQL pipeline (i.e., whether to fall back to a generic NL2SQL pipeline that does not use \name's proposed specializations). The abstention policy is based on a multi-armed bandit framework and incorporates user feedback into the decision-making process.

We evaluate \name across multiple NL2SQL benchmarks, demonstrating up to 2$\times$ improvement in execution accuracy compared to baselines that do not take advantage of past queries. Additionally, for the same execution accuracy, \name improves SQL
generation latency by 2-4$\times$ compared to the baselines. Furthermore, \name does not overfit to historical queries and exhibits robustness against shifts in query distribution. In summary, we make the following key contributions:
\begin{itemize}
    \item We present \name, a NL2SQL system that tailors itself to a given database by utilizing database-specific hints extracted from past queries to improve accuracy.
    \item We demonstrate how \name tailors its embedding and retrieval procedures to a given database, which improves SQL generation latency without sacrificing accuracy.
    \item We ensure that \name is robust to workload distribution shifts using a bandit-based policy to decide when to abstain from utilizing query hints.
    \item We present an evaluation of \name's overall performance in terms of accuracy and latency as well as microbenchmarks on its individual components.
\end{itemize}

\section{Preliminaries}
\label{sec:preliminaries}

In this section we give background on retrieval models and retrieval-augmented generation, which is useful for understanding \name.

\NewPara{Dense Retrieval Models and Document Retrieval:} Dense retrieval models are a type of information retrieval method that efficiently matches text-based queries\footnote{Here, ``queries'' is a general term that does not necessarily refer to SQL queries.} to text-based documents by encoding queries and documents into dense vectors (commonly referred to as \textit{embeddings}) in a continuous vector space, then retrieving documents whose embeddings are \textit{similar} to a given query embedding. The similarity between a query embedding $E_q$ and a document vector $E_d$ in the embedding space can be measured using various similarity metrics. One popular metric, which we use throughout this paper, is \textit{cosine similarity}, defined as:
\begin{equation}
\text{CosSim}(E_q,E_d)=\frac{E_q \cdot E_d}{\|E_q\| \cdot \|E_d\|}
\label{eq:cosine_similarity}
\end{equation}
where the numerator denotes the dot product between the query and document embeddings, and the denominator represent their respective Euclidean norms. Cosine similarity varies between -1 (denoting perfect dissimilarity) and 1 (denoting perfect similarity).

Semantically-related queries and documents should ideally have similar embeddings. A common approach for generating embeddings is to use a pretrained sentence transformer such as Sentence-BERT (SBERT)~\cite{sbert}, which has been trained on large text corpora specifically for the purpose of generating semantically meaningful embeddings of sentences such that embeddings can be compared using cosine similarity. Here, ``sentence'' refers to arbitrary pieces of text that can contain multiple English sentences. Therefore, to avoid ambiguity, for the remainder of this paper we use the term \textit{document} instead of ``sentence.''

Given a query, one can retrieve the top-K most similar documents by computing the similarity between the query embedding and all document embeddings. To enable efficient retrieval at query time, it is common to precompute document embeddings offline.

\NewPara{Retrieval-Augmented Generation (RAG)} is an approach that merges retrieval-based and generative models to improve text generation tasks.
RAG comprises three main elements: a document store, retrieval model, and generative model. The document store contains relevant text-based documents, typically along with each document's precomputed embedding. The retrieval model efficiently selects the most relevant documents based on a given query, employing techniques described above. Typically, the generative model is a pre-trained LLM, which produces contextually-relevant text based on the query and retrieved documents.

When RAG is applied to NL2SQL, the document store typically contains documents with database-specific information, such as schema information (e.g., a document describing the column names and data types of a particular table) or database documentation (e.g., a document describing the supported SQL syntax for the database's particular SQL dialect).

\section{\name System Overview}
\label{sec:system_overview}

\begin{figure*}[t]
    \begin{center}
        \includegraphics[width=\linewidth]{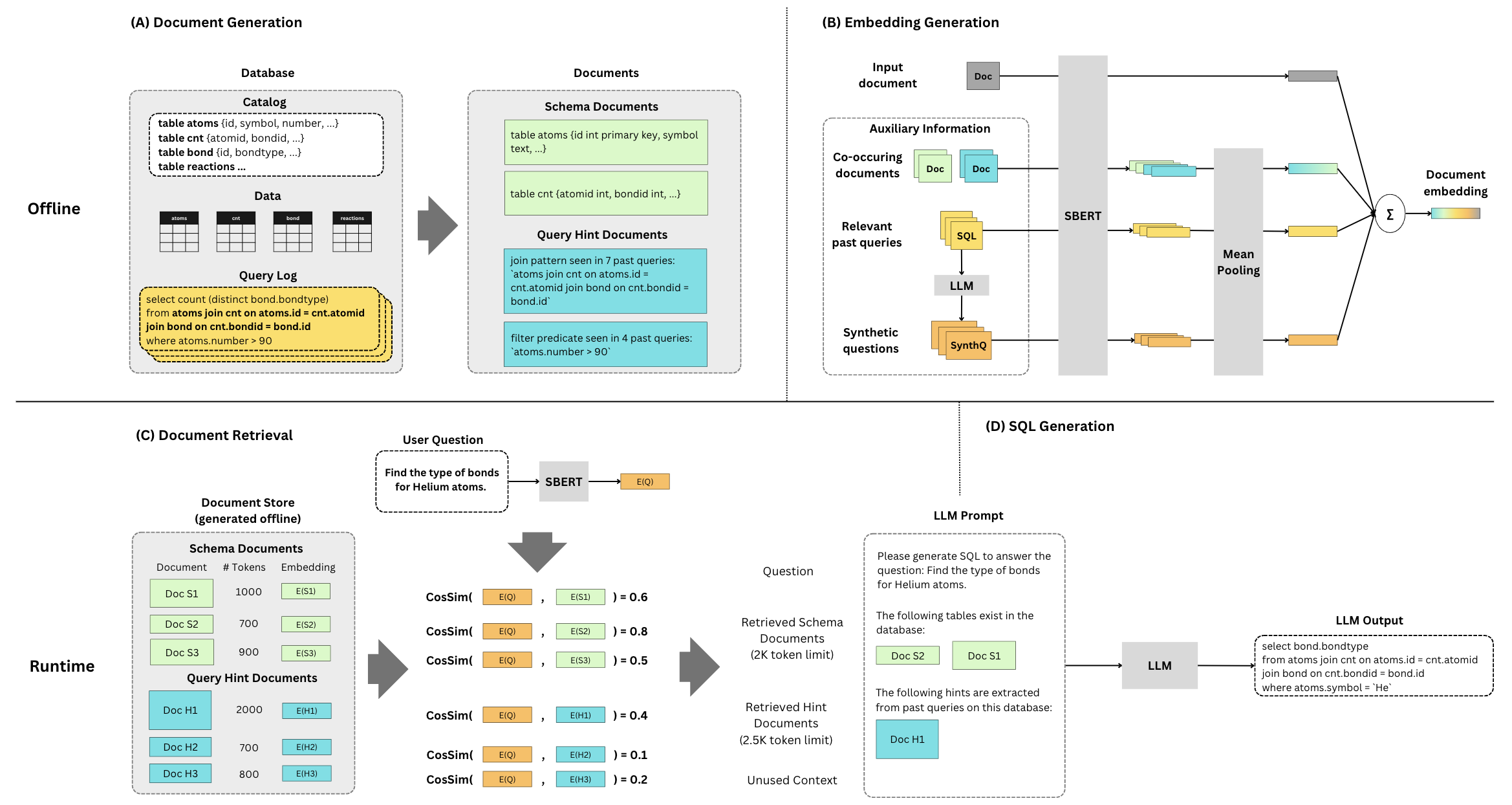}
    \end{center}
    \caption{In an offline process, \name (A) generates schema and query hint documents based on information in the database, then (B) generates an embedding for each document using a combination of the document itself and several pieces of auxiliary information related to the document. (C) At runtime, we compute similarity between the user question's embedding and document embeddings in order to retrieve the most relevant documents, up to a per-document-class token limit. (D) The prompt, which includes the question and information in retrieved documents, is used to invoke an LLM to generate the SQL query.}
    \label{fig:system_overview}
\end{figure*}

\name is a NL2SQL system which accepts a natural language question as input from the user and produces a corresponding SQL query as output. Like other recent NL2SQL systems~\cite{wang2023mac,pourreza2023dinsql,gao2023texttosql}, \name employs an LLM by creating a prompt based on the input question, invoking the LLM using the prompt, and extracting the SQL output from the LLM's response. \name's core novelty compared to prior NL2SQL systems is its ability to specialize the NL2SQL pipeline to the user's workload by taking advantage of historical query logs.

We first describe the end-to-end workflow of \name's NL2SQL pipeline, with a particular emphasis on how \name specializes each part of this pipeline using historical query logs. We then provide some further illustrative examples that intuitively showcase how historical query logs can improve NL2SQL accuracy.

\subsection{End-to-End Workflow}
\label{sec:end_to_end_workflow}
\name's NL2SQL system (shown in \cref{fig:system_overview}) is composed of an offline pipeline and an runtime pipeline. The offline pipeline preprocesses the information in the database and historical query logs so that the information can be retrieved for user questions. The offline pipeline must run before \name can start serving user questions, and it can be retriggered periodically (e.g., every night) so that \name is up to date with changes in the database schema or query workload patterns. The offline pipeline has two parts:
\begin{itemize}
    \item \textbf{Document generation:} we compile information from the database which might be useful to include in the LLM prompt into individual pieces of text called \textit{documents}. For example, prior NL2SQL techniques often generate one document for each table in the database, containing information such as the table's name and the name and data type of its columns. \name specializes to the workload by additionally generating documents containing information about common patterns in the historical query workload. \cref{sec:doc_generation} describes the process in further detail.
    \item \textbf{Document embedding generation:} each document is identified by a fixed-length vector embedding. The embedding is meant to capture semantic information about the contents of the document, so that the embedding of a document is similar (in terms of cosine similarity; see \cref{eq:cosine_similarity}) to the embeddings of user questions which make use of the document and dissimilar to the embeddings of unrelated user questions. A common approach for generating document embeddings is to feed the content of the document through a pretrained embedding model, such as SBERT~\cite{sbert}. \name specializes to the workload by generating document embeddings that are a function of not only the document's contents, but also of other related documents and related queries from the historical logs, which helps capture semantic information that is not present in the document's contents alone. \cref{sec:retrieval_overview} describes the process in further detail.
\end{itemize}
The runtime pipeline is triggered whenever the user asks a natural language question, and its output is a SQL query. The runtime pipeline has two components:
\begin{itemize}
    \item \textbf{Document retrieval:} we compute the embedding of the user question, then compute the cosine similarity between the question embedding and the document embeddings which were generated offline. Prior NL2SQL techniques would then retrieve documents in decreasing order of similarity, until the prompt limit is saturated (i.e., until the total number of tokens in all retrieved documents surpasses the token limit of the LLM prompt). However, prior techniques do not distinguish between different types of documents. \name introduces a \textit{context allocator} which performs an offline analysis to determine how much of the context to allocate for documents of each class in order to balance the precision and recall of document retrieval on the historical query workload. \cref{sec:context_allocator} describes the process in further detail.
    \item \textbf{SQL generation:} we assemble the retrieved documents into a prompt template. The same prompt template is used across all user questions. We invoke the LLM with the constructed prompt and parse the LLM response to yield the SQL query. \name does not perform any workload specialization for the SQL generation step. Indeed, prior NL2SQL systems make use of complex reasoning pipelines to perform SQL generation, e.g., by taking advantage of chain-of-thought and least-to-most prompting~\cite{pourreza2023dinsql}. These improvements to SQL generation are orthogonal and complementary to \name's use of information from past queries.
\end{itemize}
Specialized NL2SQL pipelines perform well when future questions have similar patterns as historical query logs, but can perform poorly otherwise. To avoid performance regressions, \name uses an abstention policy (\cref{sec:abstention}) to dynamically decide when to fall back to using a non-specialized NL2SQL pipeline.

\subsection{Illustrative Examples}
The example from \cref{sec:introduction} showed how past queries can help disambiguate cryptically-named tables by providing information about common join paths. We now present two further examples where past queries provide information beyond that which is already provided through the database schema.

\begin{figure}
\begin{center}
    \includegraphics[width=\linewidth]{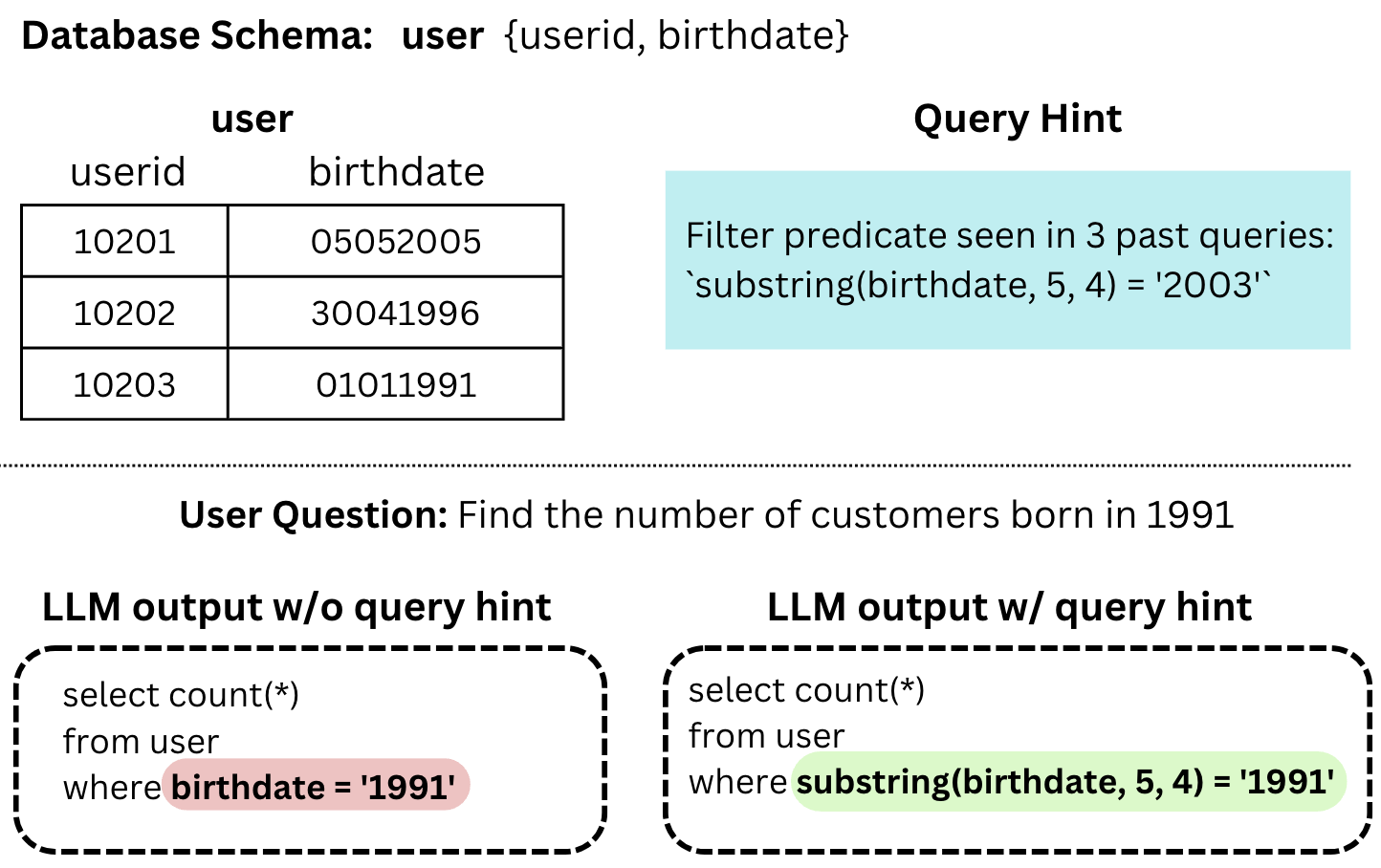}
\end{center}
\caption{Past filter predicates help interpret column values.}
\label{fig:example_filter}
\end{figure}

\NewPara{Filter Expressions:}
Filter expressions in past queries provide insights into the format of various columns, including literals within those columns, aiding the LLM in determining the appropriate filter for a query.
In \cref{fig:example_filter}, the \texttt{user} table contains user information such as ids and birth date. Notably, \texttt{birthdate} is stored as an 8-digit integer in a distinctive manner. Past queries use filter expressions of a specific form to accurately extract the year from \texttt{birthdate} values. Incorporating this query hint into the prompt assists the LLM in understanding the format of the \texttt{birthdate} column.

\begin{figure}
\begin{center}
    \includegraphics[width=\linewidth]{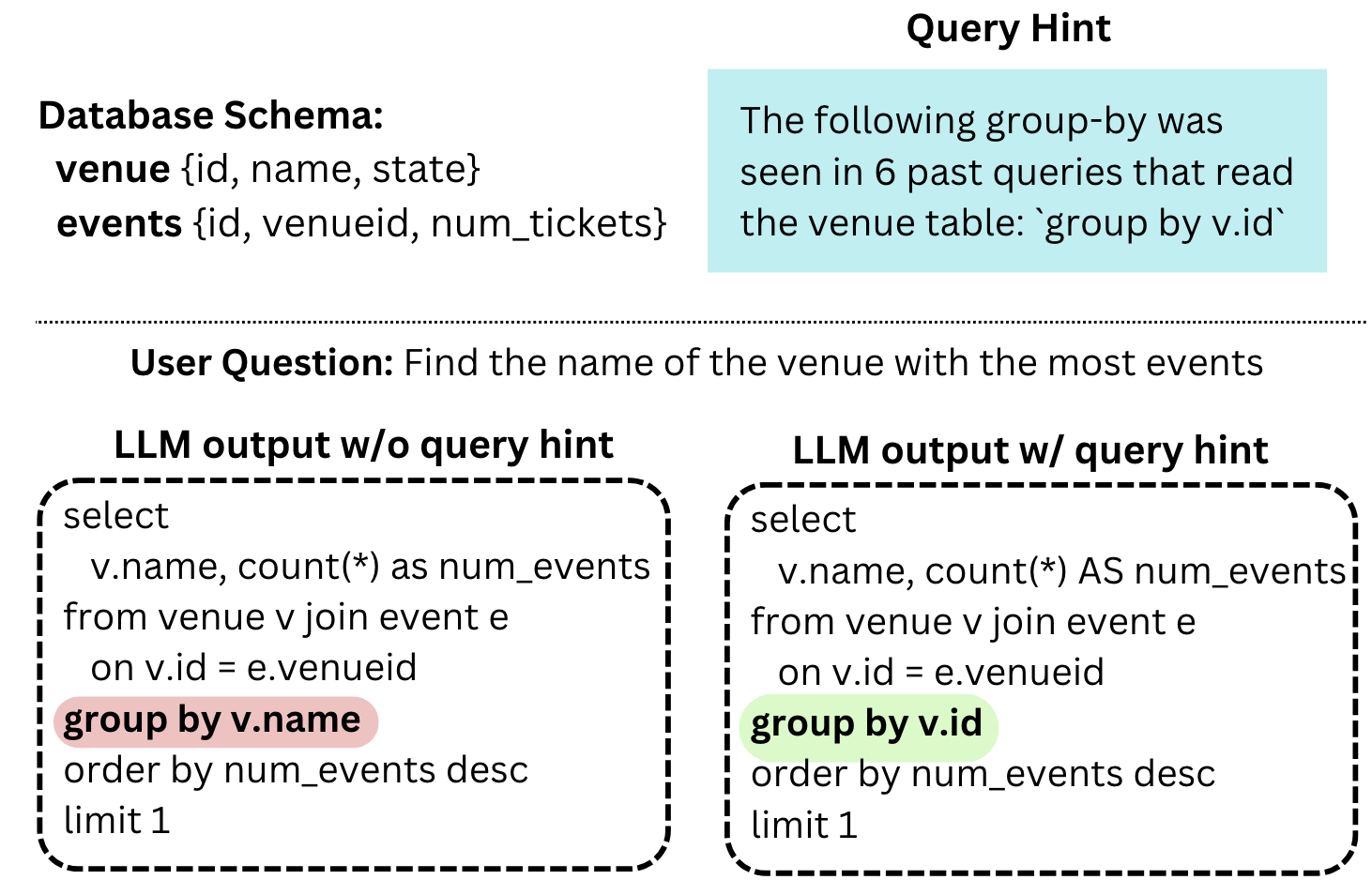}
\end{center}
\caption{Past group-by clauses help indicate an implicit primary key. Note that the SQL query on the right only executes for SQL dialects that allow bare columns in aggregation queries, like SQLite.}
\label{fig:example_group_by}
\end{figure}

\noindent\textbf{Group-By Clauses:}
In \cref{fig:example_group_by}, the \texttt{id} column serves as the primary key for the \texttt{venue} table, not the \texttt{name} column, since multiple venues may share the same name. Past aggregation queries tend to group by the \texttt{id} column, signifying that \texttt{id} refers to unique venues. If a user requests information about venues that requires aggregation, a query hint based on past group-by clauses will guide the LLM to aggregate over the \texttt{id} column instead of the \texttt{name} column.

\section{Document Store}
\label{sec:doc_generation}

\name uses information about the database schema and past query workload to help generate accurate SQL queries for user questions.
\name stores this information in two broad classes of documents: (1) schema documents, which capture information about the database schema, and (2) query hint documents, which capture information from the past query workload.

\subsection{Schema Documents}
Schema documents capture information about the database schema. Similar to prior NL2SQL techniques~\cite{pourreza2023dinsql,wang2023mac}, \name generates a document for each table in the database. By default, the schema document will contain the \texttt{CREATE TABLE} SQL statement which is used to create the table, which typically captures the following information: (1) the table's name, (2), the name, data type, default value, nullability, and other properties for each column, and (3) table properties such as the primary key and any foreign keys. All of this information helps the LLM reason about the correct SQL for a user question.

Note that the exact contents of the schema document will vary depending on the database management system. For example, some multi-node database systems define distribution keys for each table, which do not exist for single-node systems. Indeed, we found that certain additional information beyond the \texttt{CREATE TABLE} SQL statement are useful for \name, which we describe further in \cref{sec:implementation}. In general, the exact contents of the schema document are orthogonal to the core idea of \name, and \name's procedure will work with whatever schema documents are generated for a given DBMS.

\subsection{Query Hint Documents}
Unlike schema documents, which are standard in prior NL2SQL techniques, query hint documents are a novel class of documents introduced by \name. Query hint documents capture information about common query structures in the past workload. Since queries often repeat in analytic workloads, including information about past query structures should help the LLM generate accurate SQL for future user questions. There are many options for the structure of query hint documents, which vary on a tradeoff space between information content and information density:
\begin{itemize}
    \item Create a hint document for each query observed in the past workload, containing the SQL text of the query. This provides the maximum amount of information to the LLM. However, there are several disadvantages to this approach: (1) SQL queries can have very long text~\cite{tpc-not-enough,tableau-workload}, which quickly exhausts the context space. (2) SQL queries often have repetitive text, which means that hint documents would contain redundant information. For exact repeats of the same SQL query text, it is simple to deduplicate documents, but queries often do not repeat exactly but rather have repeating subcomponents (e.g., repeating filter predicates~\cite{tpc-not-enough}). (3) SQL query text often contains boilerplate text which does not provide any useful information.
    \item To alleviate the redundancy described above, an alternative is to create a hint document for each past query \textit{template}, where a template is created by removing literals from the SQL text, and then to store the literals separately. This approach might help reduce information redundancy for dashboarding and reporting workloads where the same queries are issued repeatedly with differing literals, but it does not work as well for ad-hoc query workloads where distinct templates are not as prevalent.
    \item Break down each past SQL query into subcomponents, where each subcomponent roughly corresponds to a different SQL clause, and generate a hint for each distinct subcomponent. The intuition is that queries in the past workload, despite being distinct overall, share significant subcomponents like join paths and filter predicates.
    Although this approach decreases information redundancy in documents, its disadvantage is that by extracting only a subset of information in each query, we lose some information content.
\end{itemize}
\name uses the approach of breaking down past queries into subcomponents, and in \cref{sec:implementation} we describe the specific content of hint documents used in our evaluation. However, we believe the workload-specialization techniques described in this paper are helpful regardless of the exact format of hint documents. Deciding on the content of hint documents, similar to deciding on the content of schema documents and LLM prompt engineering more generally, is more an art than a science, and further optimization to the hint document format is left to future research.

\section{Document Embedding Generation}
\label{sec:retrieval_overview}

In this section, we describe how \name generates an embedding for each of the documents in its document store. One simple workload-agnostic approach is to feed the contents of a document through an embedding model such as SBERT~\cite{sbert}, which produces an embedding that captures the semantics of the document contents. For the remainder of this section, we refer to a document's embedding produced by feeding its contents through an embedding model as a \textit{raw} document embedding. However, \name generates \textit{tailored} embeddings in a manner that makes use of the past query workload, which we show in \cref{sec:evaluation} to perform better than workload-agnostic raw SBERT embeddings.

For a given user question, we define a \textit{relevant} document as one which is helpful for answering the question. A schema document is relevant to a question if the table described in the document is used in the SQL query that answers the question. Similarly, a query hint document is relevant to a question if any content in the hint document is used in the SQL query that answers the question.

\name's embedding generation procedure takes advantage of the following intuition: the ideal document embedding is one that maximizes the similarity between the document embedding and the embeddings of future relevant user questions (and also minimizes its similarity to the embeddings of future irrelevant questions). We first describe how \name formalizes this intuition into an optimization objective which can be used to quantify the goodness of a document embedding. We then describe \name's procedure for generating document embeddings to optimize that objective.

\subsection{Optimization Objective}
We define embedding similarity as the cosine similarity (see \cref{eq:cosine_similarity}) between two embeddings. For a given pair of question embedding $\mathbf{E}_{\text{Q}}$ and document embedding $\mathbf{E}_{\text{doc}}$, we define cosine loss as:
\begin{equation*}
\text{loss}(\mathbf{E}_{\text{Q}}, \mathbf{E}_{\text{doc}}) =
\begin{cases}
1 - \cos(\mathbf{E}_{\text{Q}}, \mathbf{E}_{\text{doc}}) & \text{if doc relevant for Q} \\
\max{(0, \cos(\mathbf{E}_{\text{Q}}, \mathbf{E}_{\text{doc}})} & \text{if doc irrelevant for Q} \\
\end{cases}
\end{equation*}
Loss is high when either (1) the document is relevant to the question but the embeddings are not similar, or (2) the document is not relevant to the question but the embeddings are similar.

One challenge is that we do not know the exact questions that the user will ask in the future, so it is difficult to directly optimize for $\text{loss}(\mathbf{E}_{\text{Q}}, \mathbf{E}_{\text{doc}})$. \name tackles this challenge by taking advantage of the observation that in a stable workload, past SQL queries are likely to be similar to the queries that answer future user questions. Therefore, \name generates a synthetic question workload using the past SQL queries: for each past query, \name generates a synthetic question which, when given to a NL2SQL system, would yield that particular past query as the answer. \name employs an LLM model to generate these synthetic questions. The model is prompted with the SQL query and the schema of the tables involved in the query, and is tasked with generating a potential user question.

Given this synthetic question workload, our objective is to find an embedding for every document in order to minimize the total cosine loss for all documents in the document store $\mathbf{D}$ over all queries in the synthetic query workload $\mathbf{Q}$:
\begin{equation}
    \min \sum_{\text{doc} \in \mathbf{D}}\sum_{\text{synthQ} \in \mathbf{Q}}
    \text{loss}(\mathbf{E}_{\text{synthQ}}, \mathbf{E}_{\text{doc}})
\label{eq:optimization_objective}
\end{equation}

\subsection{Optimization Procedure}
Given this optimization objective, one possible strategy for determining document embeddings is to independently create an embedding for each document that minimizes cosine loss over the synthetic question workload. However, we found that this leads to overfitting. For example, in a degenerate case where a document is only relevant for a single synthetic question, then we would set the document embedding to be the same as the synthetic question embedding; clearly, this would not generalize to future questions which make use of the document, but which are not the same as the synthetic question.

Therefore, instead of allowing arbitrary embeddings for each document, in \name we impose a structure for document embeddings. For each document, we generate a number of \textit{proxy embeddings}, which are embeddings that should be similar to the relevant question embeddings and which capture information that may not be present in the content of the document itself:
\begin{itemize}
    \item We identify all the past queries which this document is relevant for. We use the embedding model (e.g., SBERT) to generate an embedding of each SQL query based on the query text, then average the embeddings across all queries: $\mathbf{E}_{\text{SQL}}$. We include this proxy embedding because documents are expected to be similar to the queries which use it, and SQL query information is not entirely present in documents and so would not be accounted for in the raw document embedding.
    \item For each of the relevant past queries, we prompt an LLM to generate a synthetic user question, and take the average embedding of the synthetic user questions: $\mathbf{E}_{\text{synthQ}}$. We naturally want document embeddings to be similar to the embeddings of questions which might use it.
    \item We identify co-occurring documents, which are other documents that are relevant to at least one of the queries that the current document is relevant to.
    We average the raw embeddings of all co-occurring documents: $\mathbf{E}_{\text{co-occur}}$. We include this proxy embedding because this information is useful for documents with obscure meanings, as co-occurring documents with clearer semantics can help improve their understanding. For example, in \cref{fig:main_example}, \texttt{atom} and \texttt{bond} have clearer semantics than the \texttt{cnt} table.
\end{itemize}
For a given document, its embedding is the weighted sum of each of its proxy embeddings, along with its raw embedding, $\mathbf{E}_{\text{raw}}$:
\begin{equation*}
    \mathbf{E}_{\text{doc}} = \left( w_1\cdot \mathbf{E}_{\text{raw}} + w_2\cdot \mathbf{E}_{\text{co-occur}} + w_3 \cdot \mathbf{E}_{\text{SQL}} + w_4\cdot\mathbf{E}_{\text{synthQ}} \right)
\end{equation*}
We use the same weights for all documents, i.e., the weights must be optimized once per workload, not once per document. Using the same weights for the entire workload encourages generalization.
For a given workload, we find the weights that minimize the optimization objective (\cref{eq:optimization_objective}) using gradient descent.

Proxy embeddings essentially extend the information content of the document itself. It is as if, instead of generating an embedding for a document based purely on the contents of the document itself, we are generating an embedding based on an augmented document that also includes information about relevant SQL queries and co-occuring documents.
Instead of creating proxy embeddings, we could have achieved a similar effect by generating a ``virtual'' augmented document (which includes the original document contents along with the text of co-occurring documents, relevant SQL queries and synthetic questions) and using SBERT to generate an embedding of this augmented document. However, we found that by using a weighted sum, we have more fine-grained control over the relative importance of each piece of information when constructing the embedding, which produced better embeddings.

\section{Document Retrieval}
\label{sec:context_allocator}
When the user asks a question to \name, we first convert the question into an embedding by feeding the question contents through the embedding model (e.g., SBERT). We then retrieve relevant documents from the document store and put their contents into the LLM prompt. One simple workload-agnostic approach is to retrieve documents in descending order of similarity between the question embedding and document embedding, until the LLM prompt context is filled.

However, there are several drawbacks to this simple retrieval approach, related to the existence of multiple document classes:
\begin{itemize}
    \item The optimal retrieval recall and precision may differ for each document class, where recall is defined as the fraction of relevant documents that are retrieved and precision is defined as the fraction of retrieved documents that are relevant. For example, it is intuitively more critical to have high recall for schema documents than for hint documents: if a relevant schema document is not present in the context, then \name will have difficulty generating the correct SQL because it does not know what table or column name to use, but if a relevant hint is not present in the context, \name might still generate the correct SQL.
    \item The number of documents in each class may be vastly different. By retrieving documents in a class-agnostic manner, we may end up with many more of one document class than the other, which may not be desirable.
    \item The scale of embedding similarities for each document class may be different. For example, query hint documents may naturally have embeddings that are less similar to question embeddings than schema documents, due to the difference in information content of the two documents. Some of these scale differences are mitigated due to the document embedding generation process (\cref{sec:retrieval_overview}), but there are nonetheless still effects due to the inclusion of raw document embeddings in the weighted sum that produces the tailored document embeddings.
\end{itemize}
To address these drawbacks, \name performs document retrieval in a workload-adaptive manner. \name performs an offline analysis over the past query workload to determine a \textit{context allocation} over the document classes, i.e., a way to split up the number of tokens in the context among the document classes. When performing document retrieval for a given user question, we fill the allocated context for each document class independently. That is, for each document class, we retrieve documents of that class in descending order of similarity until the class context limit is reached.

\name determines context allocation using Bayesian optimization. Specifically, we select a sample of past queries and use an LLM to generate a synthetic question for each query. The optimization objective is to identify a context allocation that maximizes \name's accuracy on the synthetic workload while adhering to a user-specified token limit. As we will describe in \cref{sec:implementation}, \name uses two classes of schema documents: table documents and column documents. Thus, the optimization is expressed as:
\begin{equation*}
    \begin{aligned}
        &\textbf{maximize} \quad &&\text{Accuracy}(t_{\text{tbl}}, t_{\text{col}}, t_{\text{hint}}) \\
        &\textbf{subject to} \quad &&0 \le t_{\text{tbl}} \le T, \quad 0 \le t_{\text{col}} \le T, \quad 0 \le t_{\text{hint}} \le T, \\
        & && t_{\text{tbl}} + t_{\text{col}} + t_{\text{hint}} \leq T
    \end{aligned}
\label{eq:accuracy_maximization}
\end{equation*}
Here, $t_{\text{tbl}}$, $t_{\text{col}}$, and $t_{\text{hint}}$ represent the number of tokens allocated to table documents, column documents, and hint documents, respectively, subject to the total token constraint $T$.
However, a naive Bayesian optimization implementation that samples configurations from $[0,T]\times[0,T]\times[0,T]$ would waste time exploring configurations that do not satisfy the token limit constraint. To address this, we reparameterize the problem by introducing variables that effectively eliminate the constraint:
\begin{itemize}
    \item $p$: Fraction of the token limit that is allocated. The remaining tokens are not allocated to any document class and remain unused.
    \item $p_{\text{tbl}}$: Fraction of the allocated tokens that is allocated to tables.
    \item $p_{\text{col}}$: Fraction of the remaining allocated tokens (after table allocation) that is allocated to columns.
\end{itemize}
Using these new variables, the token allocations are expressed as:
\begin{align*}
    t_{\text{tbl}} &= T \cdot p \cdot p_{\text{tbl}} \\
    t_{\text{col}} &= T \cdot p \cdot (1 - p_{\text{tbl}}) \cdot p_{\text{col}} \\
    t_{\text{hint}} &= T \cdot p \cdot (1 - p_{\text{tbl}}) \cdot (1 - p_{\text{col}})
\end{align*}
By reparameterizing the token allocation in this manner, the original constraint $t_{\text{tbl}} + t_{\text{col}} + t_{\text{hint}} \leq T$ is naturally satisfied, as all components are expressed as proportions of the total token limit. The reformulated optimization problem is now given by:
\begin{equation*}
    \begin{aligned}
        &\textbf{maximize} \quad &&\text{Accuracy}(p, p_{\text{tbl}}, p_{\text{col}}) \\
        &\textbf{subject to} \quad &&0 < p \le 1, \quad 0 \le p_{\text{tbl}} \le 1, \quad 0 \le p_{\text{col}} \le 1
    \end{aligned}
\label{eq:accuracy_new_maximization}
\end{equation*}
This reformulation ensures a clean optimization space, making it particularly well-suited for standard Bayesian optimization.

We use Bayesian optimization, instead of a simpler method such as gradient descent, to determine context allocation for several reasons: (1) Bayesian optimization is more sample-efficient than gradient descent (i.e., it takes fewer iterations) which is ideal for cases where evaluations of the optimization function are expensive. In our case, evaluating the objective function involves running the NL2SQL pipeline and invoking the LLM for each synthetic question, which is indeed expensive. (2) There are interactions between different document classes which makes the objective function surface complex and multi-modal. For example, increasing the context allocation for a given document class is not always desirable: although larger context improves retrieval recall, it may degrade precision, and a high concentration of irrelevant documents may in fact distract the LLM~\cite{lost_in_the_middle}. Bayesian optimization is better at finding global optima, whereas gradient descent may get stuck in local optima.

\section{Abstention Policy}
\label{sec:abstention}

\name specializes its NL2SQL workflow for a given query workload.
However, its performance may degrade when the workload characteristics change (e.g., tables which were commonly queried in the past are now used rarely), since its specializations are no longer aligned with the user questions. In this section, we describe \name's \textit{abstention policy}, which we use to decide whether user questions no longer align with the workload that \name was specialized for, and therefore to instead answer user questions using a generic, non-workload-specialized NL2SQL pipeline.

\name's abstention policy relies on the existence of two NL2SQL pipelines: \name's specialized runtime pipeline (i.e., \cref{fig:system_overview}C-D), and a generic runtime pipeline which does not use tailored embeddings or context allocations for document retrieval. Instead, the generic pipeline only retrieves schema documents by comparing similarity between the question embedding and raw document embeddings. Note that if we already ran \name's offline pipeline, there is no additional overhead for supporting a generic runtime pipeline: the schema documents and raw document embeddings needed for the generic pipeline already exist.

Conceptually, if a user question is similar to the past query workload, then we should use the specialized pipeline, and otherwise we should use the generic pipeline. Instead of directly comparing the similarity of user questions against the past query workload, \name uses runtime feedback to inform its abstention policy. We assume that after \name answers a user question with a SQL query, the user gives a binary signal (e.g., thumbs up or thumbs down) about whether they find the answer correct or useful. 

\name uses a multi-armed bandit as its abstention policy: for each incoming question, \name chooses one of the two pipelines to run. After running, we collect the binary feedback from the user. We maintain the average feedback over all past questions that are run on each pipeline, where a thumbs up maps to 1 and a thumbs down maps to 0. \name chooses which pipeline to run using an $\epsilon$-greedy strategy: with probability $\epsilon$ we choose a random pipeline, and with probability $1-\epsilon$ we choose the pipeline with the higher average historical feedback.

\name's abstention policy has two aspects which are different from a typical $\epsilon$-greedy strategy: (1) We maintain a sliding window of feedback, so that feedback that was given earlier than the window's start boundary are not considered when computing the average feedback for a pipeline. We do not want to maintain stale feedback, since we are only concerned with determining which pipeline is better for the current workload. (2) We delete all collected feedback whenever we retrigger \name's offline pipeline, i.e., whenever we regenerate documents and embeddings (see \cref{sec:end_to_end_workflow}).

\NewPara{Alternative Formulations:} Instead of a multi-armed bandit, we also considered using a contextual bandit formulation. Intuitively, the multi-armed bandit determines which pipeline the \textit{current workload} should be run on, whereas a contextual bandit determines which pipeline a \textit{given question} (i.e., the decision ``context'') should be run on. Contextual bandits may do better than a multi-armed bandit at routing questions to the best pipeline if the current workload is composed of a mix of questions that are similar and dissimilar to the past queries. However, contextual bandits require many more feedback points to learn the optimal decision strategy. We were not able to justify such a long learning process through the benchmarks in our evaluation due to the low number of questions (\cref{sec:evaluation}), though contextual bandits may provide more benefit in other benchmarks or in real-world settings.

Instead of a bandit approach, we also considered using a supervised learning approach. However, this requires a different feedback mechanism, which leads to a different user experience. To train a binary classifier that decides whether a question should be processed using the generic or specialized pipeline, we would need to first collect training data by taking each question, passing it through each pipeline to produce two different responses, and asking the user to choose the better one. This type of feedback is more informative than the simple yes/no feedback used in our bandit approach, but it places a greater mental load on users.
\section{EXPERIMENTAL EVALUATION}
\label{sec:evaluation}

We first describe the experimental setup and then present an in-depth experimental study that shows \name's performance on three NL2SQL benchmarks.
Overall, the evaluation demonstrates that:
\begin{itemize}
    \item \name achieves 10–22\% higher end-to-end SQL generation accuracy compared to other baselines
    while utilizing 2–15$\times$ smaller prompts for the same accuracy
    (\cref{sec:end_to_end_eval}).
    \item Query hint documents alone enhance performance by 4–5\%, but \name achieves greater accuracy and latency improvement through tailored embeddings and context allocation (\cref{sec:eval:ablation_study:individual_improvement}).
    \item In case of workload changes, \name's abstention policy adapts to the workload change to maintain high accuracy (\cref{sec:eval:drift_eval}).
    \item \name's workload specialization techniques are complementary to the methods in state-of-the-art NL2SQL systems and can improve their performance (\cref{sec:eval:nl2sql}).
\end{itemize}

\subsection{Experimental Setup}
\label{sec:eval:setup}
\NewPara{Datasets:} We use three NL2SQL benchmarks to test \name.
\begin{itemize}
    \item \textbf{Bird-Union} combines tables from the databases in the dev set of the BIRD benchmark~\cite{li2023llmBird} into a single database containing 71 tables and 1200 NL2SQL question-SQL pairs\footnote{We exclude primary key-foreign key (PK-FK) information from the schema and omit the evidence field provided by the benchmark to simulate real-world scenarios where this data is typically unavailable.}. By using one combined database, we simulate real-world scenarios where data is not so cleanly separated into distinct databases for every topic.
    \item \textbf{Spider-Union} combines tables from the databases in the dev set of the SPIDER benchmark~\cite{yu2019spider} into a single database containing 221 tables and 480 NL2SQL question-SQL pairs. 
    \item \textbf{FIBEN}~\cite{fiben} contains a single database containing 152 tables and 300 NL2SQL question-SQL pairs.
\end{itemize}
In general, Bird-Union and FIBEN pose a greater challenge than Spider-Union due to having more semantically intricate table names, column names, and questions.

\NewPara{Workloads:}
For each benchmark, we need to split the question-SQL pairs into two sets: one set to use for simulating historical query logs (i.e., the ``training'' set) and the other to use for simulating future user questions (i.e., the ``test'' set). We employ two types of splits:
\begin{itemize}
    \item \textbf{Random Split:} In this case, question-SQL pairs are randomly assigned to either the query log or test set with equal probability. As a result, the test set mirrors the same distribution as the query logs. This case is ideal for showcasing the benefits of \name.
    \item \textbf{Disjoint Split:} In this case, question-SQL pairs are divided in such a way that the SQL queries in the query log and test set do not access the same tables. Consequently, the test set exhibits a completely different distribution from the query logs. This case highlights workload drift when user questions have no similar counterparts in the query log. This is similar to the train/dev/test splits provided by the BIRD and SPIDER benchmarks, where the databases observed in the train set are completely disjoint from those in the dev or test sets.
\end{itemize}
All experiments have half the question-SQL pairs in the query log and the other half in the test set. Unless specified otherwise, we run experiments using Random Split.

\NewPara{Baselines:} The leading techniques on the BIRD and Spider benchmarks, such as \cite{pourreza2023dinsql,gao2023texttosql,wang2023mac,chase-sql,chess,xiyan-sql}, focus more on obtaining the best answer using superior LLMs or better decomposition or prompting techniques (e.g., least-to-most prompting).
On the other hand, the goal of our evaluation is to assess how incorporating information from past query logs influences NL2SQL performance. The techniques described in this paper are orthogonal and complementary to the techniques on the benchmark leaderboards and can be combined for further improvement in accuracy (see \cref{sec:eval:nl2sql}).
Therefore, to isolate the performance effects of using query log information and avoid confounding factors such as LLM model and prompting techniques, in our evaluation we adhere to using the same LLM (Claude 3 Haiku 1.2~\cite{claude_instant}) and perform SQL generation via a single call to the LLM for \name and for all baselines.

We use two baselines:
\begin{itemize}
    \item \textbf{SBERT:} In this baseline, we use SBERT~\cite{sbert} (specifically, the \texttt{all-MiniLM-L6-v2} model~\cite{SBERT-implementation}) to generate embeddings for user questions and for documents. This baseline does not use any workload-related specializations, i.e., it does not store query hint documents, does not generate workload-tailored embeddings, and does not perform an offline analysis for prompt context allocation. For a given user question, this baseline retrieves documents into the LLM prompt in order of decreasing embedding similarity until hitting a specified prompt token limit.
    \item \textbf{BM25:} In this baseline, we use BM25~\cite{bm25}, a lexical retrieval method, to retrieve documents. BM25 is often used for retrieval over long documents, where SBERT-based models might not perform as well. All other aspects of the baseline are the same as \textbf{SBERT}.
\end{itemize}

\NewPara{Metrics:} We evaluate the accuracy of each NL2SQL approach based on execution accuracy (EX), which measures the fraction of question-SQL pairs for which the execution results of the NL2SQL-generated SQL query and the ground-truth SQL query matches. Following the precedent of earlier papers, we measure improvement in accuracy in absolute numbers instead of relative numbers. For example, if execution accuracy increases from 20\% to 60\%, we refer to a 40\% improvement in accuracy instead of a 3$\times$ improvement. We also evaluate the latency of each NL2SQL approach, which includes the latency of retrieving relevant documents to create an LLM prompt and the latency of invoking the LLM using the prompt.
We repeat each experiment 5 times and we report the median value of each metric being measured in the experiment.

\subsection{Implementation Details}
\label{sec:implementation}
\name uses SBERT as its embedding model for generating question embeddings and raw document embeddings.

As described in \cref{sec:doc_generation}, \name's fundamental contributions are not the exact format of documents. Here, we describe the document format we used for our evaluation, but alternative formats may work better for other workloads.

\subsubsection{Schema Documents}
\name uses two classes of schema documents: table documents and column documents. Each column document contains information specific to a given column, including the column name, table name, and the ten most commonly-occurring column values. We separated table documents from column documents because we found that including samples for every column in a unified schema document results in very long documents which quickly use the limited context space in the LLM's prompt. Furthermore, if a question only requires a subset of columns from a table, it is unnecessary to retrieve the entire table document.

\subsubsection{Query Hint Documents}
\name generates the following types of query hints from past queries:
\begin{itemize}
    \item Join path hints: for each query, we extract the set of tables which are scanned and the join conditions between the tables.
    \item Filter hints: for each query, we construct a document for each filter. This includes the names of the tables whose columns are referenced in that filter.
    \item Group-by hints: we construct one document for the entire group-by condition, including the names of the tables whose columns are referenced in the group-by.
\end{itemize}
These query hints cover the types of SQL clauses which we most commonly observed in our benchmarks and which appeared most useful to \name, but they are not exhaustive. We believe that \name can be easily extended to generate hints for other SQL clauses if needed.

If a query hint is found in multiple past queries, we merge the query hint documents and add a counter for the number of times this hint has been observed in the past.

\begin{figure*}[t]
    \begin{center}
        \includegraphics[width=0.9\linewidth]{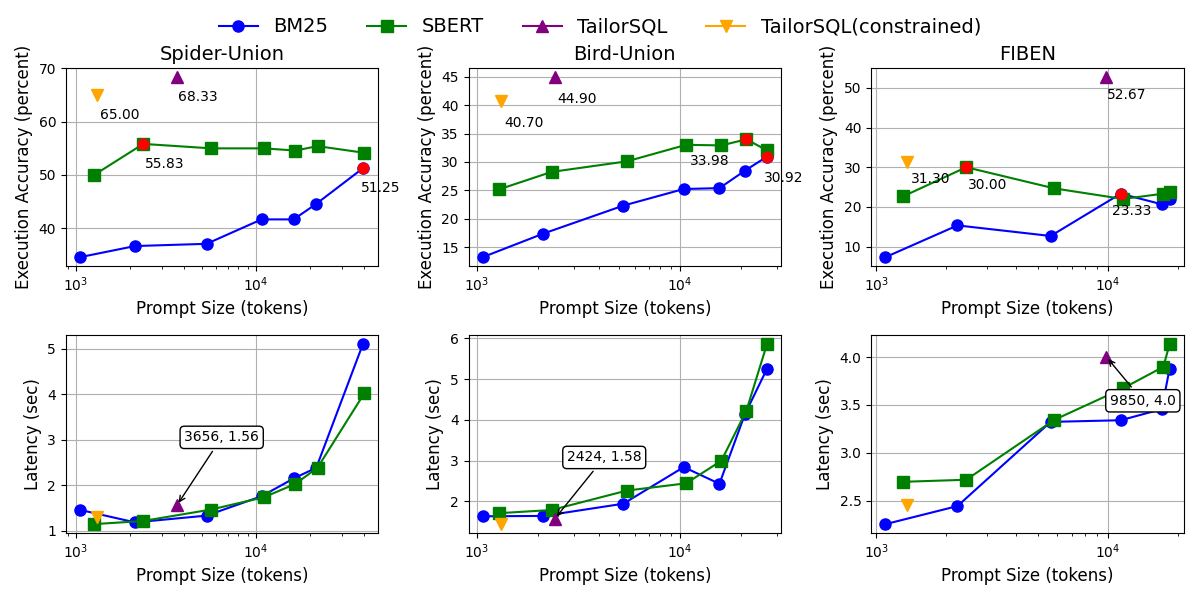}
    \end{center}
    \caption{\name achieves 12.5\%, 10.9\% and 22.7\% higher match execution accuracy compared to the next best baseline (SBERT). \name achieves the greatest accuracy gains on FIBEN due to the benchmark's complex queries and cryptically-named schema items, which benefit from query hints. Additionally, to achieve the same accuracy, \name (constrained) uses 2-15$\times$ times fewer tokens and incurs lower latency than the baselines.}
    \label{fig:expt_1}
\end{figure*}

\subsection{End-to-End Evaluation}
\label{sec:end_to_end_eval}

The objective of this experiment is to demonstrate that \name can enhance performance by leveraging past queries. In \cref{fig:expt_1}, the first row shows the SQL execution accuracy of \name and the baselines across three benchmarks. 
Additionally, we include a constrained \name baseline, where we artificially limit the context size for documents to 1K tokens\footnote{We guarantee that the Bayesian optimization algorithm will select a context allocation with at most 1K tokens by down-scaling all candidate allocations to fit within the limit.}, in order to evaluate how \name performs in terms of accuracy if the user requires low latency.
For the baselines, we vary the limit on the LLM prompt context size. For \name, only a single point is shown since the context allocator automatically determines the context size for each document class.

Accuracy for all techniques varies across benchmarks, reflecting the varying levels of benchmark difficulty. \name consistently achieves the highest accuracy across all benchmarks compared to the baselines, while BM25 consistently underperforms. Specifically, \name achieves 12.5\%, 10.9\%, and 22.7\% higher accuracy than the next best baseline on the three benchmarks, respectively.
\name achieves more significant accuracy improvements on FIBEN than the other benchmarks due to FIBEN's complex queries (e.g., nested queries, multi-table joins) and cryptic table names, which especially benefit from query hints and tailored embeddings.

Among the two baselines, the BM25 baseline typically requires more prompt tokens than the SBERT baseline to achieve its best accuracy. This is because BM25 is worse at document retrieval than SBERT and therefore requires more prompt tokens to achieve a sufficiently high document recall to generate accurate SQL queries. Note that for SBERT, increasing the prompt size only improves accuracy up to a certain point, after which further increases in prompt size cause accuracy to drop. This implies that as prompt size increases past the optimal point, the additional documents retrieved into the prompt have an increasing likelihood of being irrelevant, which degrades retrieval precision more than it improves recall. This observation is consistent with findings that adding irrelevant information to LLMs can reduce accuracy~\cite{shi2023large}.

The SBERT baseline achieves its best accuracy with 2.3K, 21.5K, and 2.5K prompt tokens for the Spider-Union, Bird-Union, and FIBEN benchmarks, respectively. In contrast, \name matches or exceeds SBERT’s best accuracy using at most 1.3K tokens\footnote{This is larger than the context allocation limit of 1K tokens because total prompt size also includes the question and other boilerplate.} for the same benchmarks, as shown by the constrained \name baseline. As a result, \name achieves the same accuracy as SBERT while using 1.8$\times$, 15$\times$, and 1.9$\times$ fewer tokens.

The second row of \cref{fig:expt_1} depicts the latency of invoking the NL2SQL pipeline for a user question, which includes both document retrieval and SQL generation.
Latency is primarily influenced by the latency of the large language model (LLM) calls, which depend on factors such as the number of input and output tokens. Generally, latency increases as the number of input prompt tokens grows.
Compared to the SBERT baseline, \name incurs higher latency because it uses more input prompt tokens than the SBERT baseline, since \name includes both schema and hint documents in its prompts, whereas the SBERT baseline only includes schema documents, though this effect is mitigated using \name's tailored embeddings for more effective document retrieval. However, if the user is concerned with latency, they can constrain \name to use a smaller prompt size, which achieves much lower latency than the SBERT baseline.

\subsection{Ablation Study}
\label{sec:eval:ablation_study}
We perform an ablation study for the components of \name's end-to-end workflow, as well as for the components of its workload-tailored embeddings.

\subsubsection{End-to-End Workflow}
\label{sec:eval:ablation_study:individual_improvement}
\begin{table}[]
\small
\begin{tabular}{|l|r|r|r|}
\hline
\textbf{}              & \multicolumn{1}{l|}{\textbf{Accuracy}} & \multicolumn{1}{l|}{\textbf{Latency (s)}} & \multicolumn{1}{l|}{\textbf{Prompt Tokens}} \\ \hline
TailorSQL              & 0.449                                  & 1.578                                  & 2424                                      \\ \hline
w/o Tailored Embeddings & 0.409                                  & 3.34                                  & 5999                                     \\ \hline
w/o Context Allocator      & 0.369                                  & 6.29                                 & 41975                                     \\ \hline
w/o Query Hints        & 0.382                                  & 3.56                                & 6384                                      \\ \hline
\end{tabular}
\caption{\name Ablation Study on Bird-Union Benchmark}
\label{eval:ablation_study}
\end{table}

In this section, we analyze the performance impact of each key technique employed by \name.
\cref{eval:ablation_study} presents the impact of disabling each of \name's components, when evaluated on accuracy on Bird-Union. Disabling tailored embeddings results in a noticeable increase in context size and a corresponding rise in latency. This is because less effective embeddings lead to a larger context being required to gather all the necessary information, resulting in an accuracy drop due to the reduced precision in identifying relevant documents. When the context allocator is disabled, a large fixed context is used to maintain high recall. However, this results in high latency (since LLM invocation latency is correlated with the size of the input prompt) and reduced precision when retrieving relevant documents, negatively impacting accuracy. Lastly, disabling query hints causes an increase in context usage and a reduction in accuracy. Without query hints, the system lacks crucial information that assists in SQL generation, forcing the retrieval of more schema documents, which contributes to the higher context size. In summary, each component plays a critical role in maintaining optimal accuracy and latency for \name.

\subsubsection{Tailored Embeddings}
\label{sec:eval:ablation_study:embedding_model}

\begin{figure}
    \begin{center}
        \includegraphics[width=\linewidth]{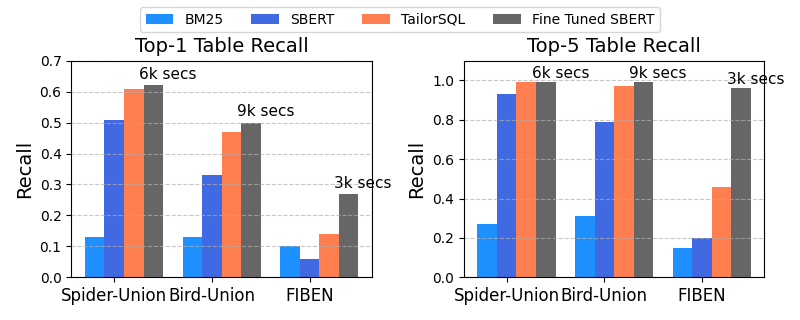}
    \end{center}
    \caption{\name's tailored document embeddings achieve higher top-1 and top-5 table document recall compared to baselines. Fine-tuned SBERT achieves better recall than \name but requires higher training time (shown above each bar).}
    \label{fig:expt_3}
\end{figure}

\begin{figure}
    \begin{center}
        \includegraphics[width=\linewidth]{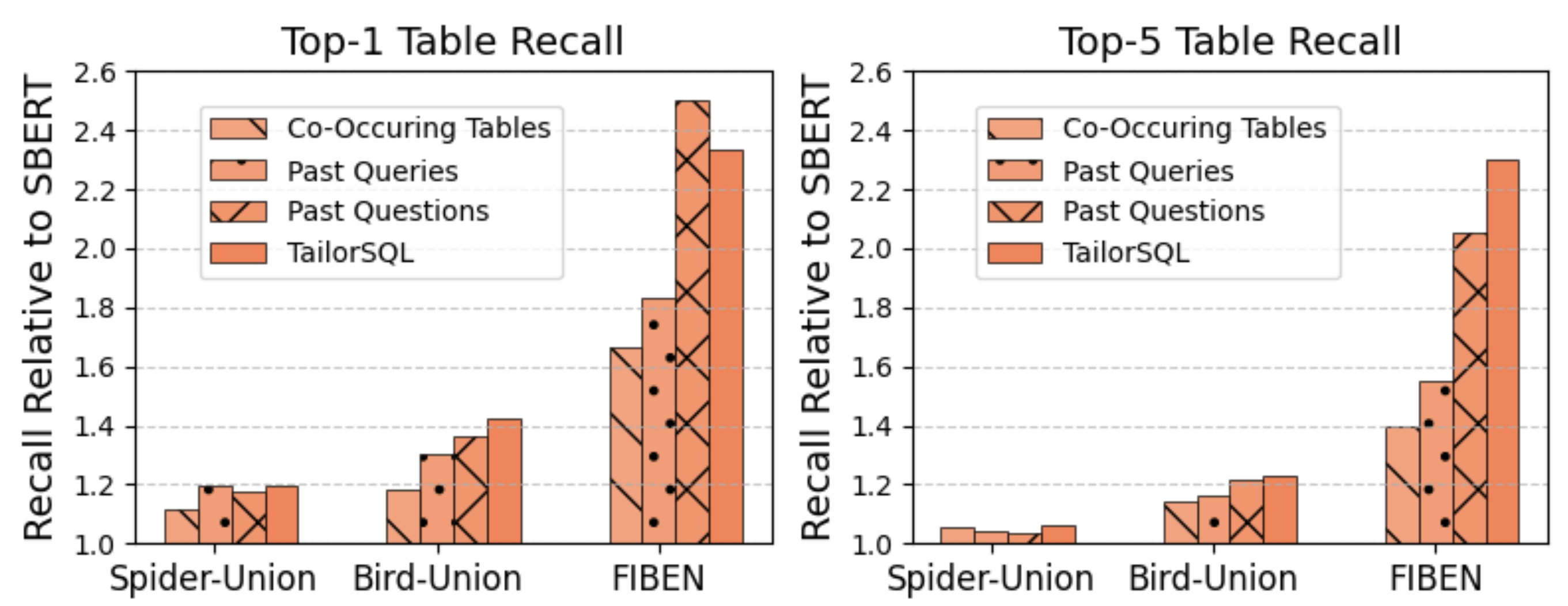}
    \end{center}
    \caption{Improvement in recall achieved by including each individual factor into \name's tailored embeddings, while ignoring other factors. Recall improves the most when all factors are used (denoted by the bar marked ``TailorSQL'').}
    \label{fig:expt_5}
\end{figure}

This section evaluates the effectiveness of \name's tailored embeddings for documents (see \cref{sec:retrieval_overview}). The first takeaway that \name's tailored embeddings perform better than using SBERT and BM25 to generate document embeddings. In Fig.~\ref{fig:expt_3}, each bar represents the table recall for strategies selecting the top-K relevant tables for a user question. Top-K document recall measures the fraction of user questions where the top-K most similar documents include all relevant documents. \name consistently outperforms vanilla SBERT embeddings by specializing document embeddings based on the past query workload, leading to significant accuracy improvements (\cref{sec:end_to_end_eval}). BM25 performs the worst, consistent with its low accuracy in \cref{fig:expt_1}.

The second takeaway is that \name's tailored embeddings achieve worse recall than the approach of fine-tuning SBERT and using the fine-tuned model to generate embeddings. However, fine-tuning requires up to 2.5 hours of additional training time (as shown in \cref{fig:expt_3}), whereas \name's tailored embeddings require less than one minute of training on a CPU. Furthermore, fine-tuned models require additional storage space and may not generalize as well as \name's purposefully underparameterized model. While fine-tuned embedding models is a great approach when maximum recall is desired, we decide to use \name's method of learned embeddings to minimize training time and storage overhead.

We now explore the effect of each factor that contributes to \name's tailored document embeddings.
\cref{fig:expt_5} illustrates the impact on recall when integrating each of the three factors individually into the tailored embedding, along with raw SBERT document embeddings, while ignoring the two other factors.
Among the three factors, embeddings based on synthetic past questions demonstrate the most substantial improvement across all benchmarks, while the incorporation of co-occurring documents contributes the least improvement.

\subsection{Robustness against Workload Drift}
\label{sec:eval:drift_eval}

\begin{figure}
    \begin{center}
        \includegraphics[width=\linewidth]{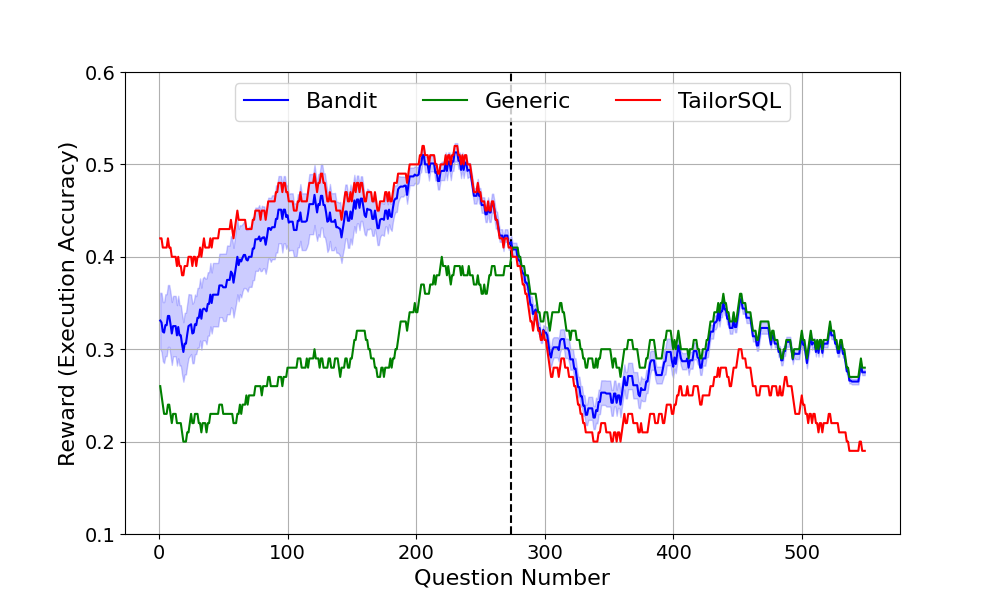}
    \end{center}
    \caption{All questions before the vertical dotted line are similar to past queries, while all questions afterwards are dissimilar to past queries. \name's bandit-based abstention policy is able to select the better pipeline in case of this workload shift, and performs better than purely using \name's workload-tailored pipeline or a generic SBERT-based pipeline on all questions.}
    \label{fig:expt_bandit}
\end{figure}

In this section, we evaluate whether \name's abstention policy is able to correctly select whether user questions should run on \name's workload-tailored pipeline or on a generic NL2SQL pipeline. Intuitively, we expect the policy to choose to run on \name's pipeline if the user question follows similar patterns as the past query workload, and to choose to run the generic pipeline otherwise. In particular, as the user question characteristics drift over time, we expect the policy to dynamically switch from favoring \name's pipeline to favoring a generic pipeline.

\cref{fig:expt_bandit} illustrates the behavior of the abstention policy under workload drift, which occurs over the course of many user questions. The first half of user questions follows the Random Split workload (see \cref{sec:eval:setup}), which means that the user questions are similar to the past queries; the second half of user questions follows the Disjoint Split workload, which means that the user questions are dissimilar to the past queries. The plot shows the reward obtained by the abstention policy, which represents the execution match accuracy over a sliding window of 100 questions, as user questions are submitted. A 95\% confidence interval is plotted around the abstention policy's reward to indicate variability. For comparison, we include the performance of only using \name's pipeline for every question and of only using a generic pipeline, equivalent to the SBERT baseline.

As expected, a policy of only using \name's pipeline performs better during the first half of the workload, while the policy of only using the generic SBERT baseline performs better in the second half. Initially, the abstention policy's bandit algorithm explores both pipelines, quickly converging to \name's pipeline in the first half. Upon detecting the distribution shift at the midpoint, the accuracy briefly drops, but the bandit algorithm soon adapts, selecting the SBERT pipeline as the optimal choice for the second half. Thus, the bandit-based abstention policy ensures adaptation to workload drift.

\subsection{Impact on other NL2SQL methods}
\label{sec:eval:nl2sql}
We now present empirical evidence that leveraging past queries to improve NL2SQL generation is beneficial for existing NL2SQL systems. State-of-the-art NL2SQL methods, such as those on the Spider and BIRD benchmarks, focus on enhancing SQL generation through advanced reasoning techniques like Chain-of-Thought~\cite{wei2023chainofthought} and Self-Consistency~\cite{wang2023selfconsistencyimproveschainthought}. The core innovation of \name lies in its use of past queries, which is complementary and orthogonal to these techniques, allowing \name to augment their performance.

To demonstrate this, we modified two well-known NL2SQL systems, DIN-SQL~\cite{pourreza2023dinsql} and MAC-SQL~\cite{wang2023mac}, to incorporate \name's provided prompt as the initial prompt, and compared this against the use of SBERT-based retrieval for the initial prompt. \cref{eval:bird-union_nl2sql,eval:spider-union_nl2sql} show that using \name as the initial prompt significantly improves accuracy and reduces SQL generation latency compared to SBERT-based retrieval across both benchmarks and NL2SQL systems.
Note that the accuracy of DIN-SQL and MAC-SQL in our results differs from the values reported on the public Spider and BIRD leaderboards due to modifications in our setup---we introduced changes to the benchmarks by combining each benchmark's databases into one Union database and excluding evidence from the BIRD benchmark, and we employed a different LLM than the one originally used to tune the DIN-SQL and MAC-SQL prompts.

\begin{table}[]
\small
\begin{tabular}{|l|r|r|r|}
\hline
\textbf{Baseline} & \multicolumn{1}{l|}{\textbf{\begin{tabular}[c]{@{}l@{}} Accuracy\end{tabular}}} & \multicolumn{1}{l|}{\textbf{Latency (s)}} & \multicolumn{1}{l|}{\textbf{Prompt Tokens}} \\ \hline
DIN-SQL+SBERT      & 0.5                                                                                              & 61.79                                          & 40255                                     \\ \hline
DIN-SQL+TailorSQL  & 0.6125                                                                                           & 15                                             & 1460                                      \\ \hline
MAC-SQL+SBERT      & 0.516                                                                                            & 38.16                                          & 40212                                     \\ \hline
MAC-SQL+TailorSQL  & 0.654                                                                                            & 3.419                                          & 1397                                      \\ \hline
\end{tabular}
\caption{(Spider-Union) \name improves state-of-the-art NL2SQL systems (DIN-SQL and MAC-SQL) by using past queries.}
\label{eval:spider-union_nl2sql}
\end{table}

\begin{table}[]
\small
\begin{tabular}{|l|r|r|r|}
\hline
\textbf{Baseline} & \multicolumn{1}{l|}{\textbf{\begin{tabular}[c]{@{}l@{}}Accuracy\end{tabular}}} & \multicolumn{1}{l|}{\textbf{Latency (s)}} & \multicolumn{1}{l|}{\textbf{Prompt Tokens}} \\ \hline
DIN-SQL+SBERT      & 0.219                                                                                            & 158                                            & 26899                                     \\ \hline
DIN-SQL+TailorSQL  & 0.4                                                                                              & 21                                             & 1889                                      \\ \hline
MAC-SQL+SBERT      & 0.255                                                                                            & 34.73                                          & 26856                                     \\ \hline
MAC-SQL+TailorSQL  & 0.47                                                                                             & 4.03                                           & 1851                                      \\ \hline
\end{tabular}
\caption{(Bird-Union) \name improves state-of-the-art NL2SQL systems (DIN-SQL and MAC-SQL) by using past queries.}
\label{eval:bird-union_nl2sql}
\end{table}
\section{Related Work}

\NewPara{LLMs for NL2SQL:}
Today, the leaderboards for NL2SQL benchmarks like Spider and BIRD are dominated by LLM-based solutions~\cite{nl2sql-survey,chase-sql,chess,xiyan-sql}, while earlier methods leading the benchmarks were mostly based on manually-tweaked encoder-decoder LSTM-based architectures, e.g.~\cite{scholak2021picard,wang2021ratsql}. Top-performing NL2SQL systems primarily focus on question representation and information organization. DIN-SQL~\cite{pourreza2023dinsql} uses decompositions and intermediate query representations following chain-of-thought~\cite{wei2023chainofthought} and least-to-most prompting~\cite{zhou2023leasttomost} paradigms. DAIL-SQL~\cite{gao2023texttosql} evaluates different methods of question representations, code representation, information (metadata) organization and picks the best combination of the three. Contrary to our approach that focuses on optimizing in-domain performance, DAIL-SQL focuses on cross-domain in-context learning and specifically masks domain-specific keywords. CodeS~\cite{codes} is a pretrained LLM designed specifically for NL2SQL.
These methods are orthogonal to our idea of incorporating past query history and combining these method into \name can further improve performance. SNAILS~\cite{snails} shows that NL2SQL techniques generally perform worse on databases that use less natural schema names, which further motivates the need for NL2SQL techniques like \name that can understand obscurely-named tables and columns.

\NewPara{Retrieval Methods:}
Retrieval-augmented generation (RAG)~\cite{lewis2021retrievalaugmented} has been employed to boost LLM accuracy across various NLP domains.
\textit{Bi-encoding} retrieval methods encode both question and document separately using the same embedding transformation, and compute their similarity via a distance metric such as cosine similarity.
Recent literature shows that variants of the BERT model for embedding~\cite{devlin2019bert,MacAvaney_2019} exhibit the best retrieval accuracy. However, BERT models are expensive to fine-tune due to their large capacity and a notorious scarcity of training data across many domains. Therefore, a pragmatic approach is to rely on bi-encoded similarity using pretrained models.
\textit{Cross-encoding}~\cite{akkalyoncu-yilmaz-etal-2019-cross,nogueira2020passage} feeds the concatenated question-document pair into BERT and trains a FCN classifier layer on its vector representation. It commonly outperforms bi-encoded similarity scoring since it is able to capture more complex cross-feature interactions between the question and document. However, this approach is often prohibitively expensive at inference time, as it requires full encoding of each question-document pair at retrieval time.

\NewPara{Fine-tuning LLMs:} Fine-tuning involves adjusting a pretrained model on a specific, often narrower, dataset or task to enhance its performance in that particular domain. Fine-tuning techniques are commonly classified into supervised~\cite{ouyang2022training,wang2022super,taori2023alpaca}, unsupervised~\cite{zhou2023lima}, and reinforcement learning~\cite{touvron2023llama,rafailov2023direct} based methods. In the case of NL2SQL pipelines, the past query workload along with synthetically generated user questions could be used as input-output pairs for supervised fine-tuning of the model. This method could potentially deliver better results than the RAG-style solution.
However, there are several practical downsides to fine-tuning.
First, users may have privacy concerns about using their data to train LLMs shared across users.
Second, training and maintaining a separate fine-tuned LLM per database is an expensive operation, especially given the dynamic nature of databases where the data and query workload frequently change.

\section{Conclusion}

We introduced \name, an NL2SQL system that tailors itself to a specific database by leveraging the database's query history to enhance document generation, document embedding, and document retrieval in a RAG-based NL2SQL pipeline. We further introduced a bandit-based abstention policy which dynamically determines when \name's workload-specialized pipeline is no longer suitable for the current workload, thereby avoiding performance degradations due to workload drift. \name demonstrates consistent accuracy and latency improvements across three NL2SQL benchmarks.

\bibliographystyle{ACM-Reference-Format}
\bibliography{text2sql}

\end{document}